\documentstyle[aas2pp4]{article}

\begin{document}

\title{Ultraviolet Signposts of Resonant Dynamics \\
in the Starburst-Ringed Sab Galaxy, M94 (NGC 4736)}
\author{William H. Waller\altaffilmark{1}\altaffilmark{2}, 
Michael N. Fanelli\altaffilmark{3}\altaffilmark{2},
William C. Keel\altaffilmark{4},
Ralph Bohlin\altaffilmark{5}, 
Nicholas R. Collins\altaffilmark{2}, 
Barry F. Madore\altaffilmark{6},
Pamela M. Marcum\altaffilmark{7},
Susan G. Neff\altaffilmark{8}, 
Robert W. O'Connell\altaffilmark{9}, 
Joel D. Offenberg\altaffilmark{2}, 
Morton S. Roberts\altaffilmark{10}, 
Andrew M. Smith\altaffilmark{8}, 
and Theodore P. Stecher\altaffilmark{8}}

\altaffiltext{1}{Tufts University, Department of Physics and Astronomy, 
Medford, MA  02155}

\altaffiltext{2}{Raytheon ITSS Corporation, NASA Goddard Space Flight Center,
Laboratory for Astronomy and Solar Physics, Code 681, Greenbelt, MD  20771}

\altaffiltext{3}{University of North Texas, Department of Physics, Denton,
TX 76203}

\altaffiltext{4}{University of Alabama, Department of Physics and Astronomy, 
P. O. Box 870324, Tuscaloosa, AL  35487-0324}

\altaffiltext{5}{STScI, Homewood Campus, Baltimore, MD  21218}

\altaffiltext{6}{Infrared Processing and Analysis Center, Caltech, M/S
100-22, 770 So. Wilson Ave., Pasadena, CA  91125}

\altaffiltext{7}{Texas Christian University, Department of Physics, Box 298840,
Fort Worth, TX  76129}

\altaffiltext{8}{NASA Goddard Space Flight Center, Laboratory for Astronomy
and Solar Physics, Code 680, Greenbelt, MD  20771}

\altaffiltext{9}{University of Virginia, Department of Astronomy, P. O. Box
3818, Charlottesville, VA  22903}

\altaffiltext{10}{National Radio Astronomy Observatory, 520 Edgemont Rd.,
Charlottesville, VA  22903--2475}

\begin{abstract}

The dynamic orchestration of starbirth activity in the starburst-ringed 
galaxy
M94 (NGC 4736) is investigated using images from 
the Ultraviolet Imaging Telescope (FUV-band),
Hubble Space Telescope (NUV-band), 
Kitt Peak
0.9-m telescope (H$\alpha$, R, and I bands), and Palomar 
5-m telescope (B-band), along with spectra from the International Ultraviolet 
Explorer and Lick 1-m telescopes.
The wide-field UIT image shows FUV emission from (a) 
an elongated nucleus, 
(b) a diffuse inner disk, where H$\alpha$ is observed in {\it absorption},
(c) a bright inner ring of H II regions at the perimeter of the inner disk 
(R = 48$''$ = 1.1 kpc), 
and (d) two 500-pc size knots of hot stars exterior to the ring on 
diametrically opposite sides of the nucleus (R = 130$''$ = 2.9 kpc).
The HST/FOC image resolves the NUV
emission from the nuclear region into a bright core and a faint
20$''$-long ``mini-bar'' at
a position angle
of 30 deg.
Optical and IUE spectroscopy of the nucleus and 
diffuse inner disk indicates a 
$\sim$10$^{7-8}$ yr-old stellar population from low-level 
starbirth activity blended with some LINER activity.
Analysis of the H$\alpha$, FUV, NUV, B, R, and I-band emission along
with other observed tracers of stars and gas in M94 indicates that most of the 
star
formation is being orchestrated via ring-bar dynamics involving 
the nuclear 
mini-bar, inner ring, oval disk, and outer ring.  The inner 
starburst ring and 
bi-symmetric knots at intermediate radius, 
in particular, argue for bar-mediated resonances 
as the primary drivers of evolution in M94 at the present epoch.  
Similar processes may be governing 
the evolution of the ``core-dominated'' galaxies that have been observed 
at high redshift.  The gravitationally-lensed ``Pretzel Galaxy'' (0024+1654)
at a redshift of $\sim$1.5 provides an important precedent in this regard.
\end{abstract}


\section{Introduction}
Star-forming rings or ``pseudorings'' are common to early and 
intermediate-type spiral galaxies (cf.\markcite{Athan85}
Athanassoula \& Bosma 1985; \markcite{buta95b}Buta, Purcell, \& Crocker 1995; 
\markcite{bc96}Buta \& Combes 1996), 
including our own Milky Way galaxy (cf. \markcite{gm82}Gusten \& Mezger 1982; 
\markcite{Clemens88}Clemens, Sanders, \& Scoville  
1988; \markcite{Waller90a}Waller 1990a).  
Such ring-like accumulations of gas and 
associated starbirth activity may have helped to build the 
inner parts of many primeval disk galaxies 
(\markcite{Friedli95}Friedli \& Benz 1995), as 
exemplified by the recent discovery 
of a starburst-ringed 
galaxy at $z \sim 1.5$ (\markcite{Colley96}Colley, Tyson, \& Turner 1996; 
\markcite{Tyson97}Tyson et al. 1997).  The formation and 
maintenance of 
these starburst rings are often attributed to orbital resonances with 
rotating bar or ``oval'' asymmetries in the stellar disks (cf. 
\markcite{Combes94}Combes 1994, 
\markcite{Byrd94}Byrd et al. 1994; \markcite{Combes95}Combes et al. 1995; 
\markcite{bc96}Buta \& Combes 1996 and references therein).  
However, other dynamical mechanisms --- including gravitational 
instabilities 
(\markcite{Elmegreen92}Elmegreen 1992, \markcite{Elmegreen94}1994; 
\markcite{Kenney97}Kenney \& Jogee 1997), 
outward propagating star formation (\markcite{Walker88}Walker, Lebofsky, \&
Rieke 1988; 
\markcite{Waller92}Waller, Gurwell, \& Tamura 1992), 
and even 
radially-driven pileups from nuclear outbursts (Waller et al. 1992; 
\markcite{Tenorio97}Tenorio-Tagle et al. 1997) --- may play 
significant roles in 
orchestrating some of the starburst rings that are observed.
\footnote{Collisionally-induced ``ring galaxies'' such as the Cartwheel Galaxy 
are thought to be morphologically and dynamically 
distinct from the more common ``ringed galaxies'' considered herein (cf. 
\markcite{Athan85}Athanassoula \& Bosma 1985; \markcite{Marcum92}
Marcum et al. 1992).}

As the closest early-type spiral galaxy of low inclination, M94 (NGC 4736) has 
received concentrated attention from both observers and theorists.  This 
(R)SA(r)ab-type galaxy (\markcite{rc3}de Vaucouleurs et al. 1991) 
is noted for its inner ring of 
ongoing starburst activity (R $\approx$ 45$''$), 
oval stellar distribution at intermediate radius 
(R $\approx$ 220$''$) (cf. \markcite{Mulder93}Mulder \& van Driel 
1993; \markcite{Mulder95}Mulder 1995; \markcite{Mollenhoff95}
Mollenhoff, Matthias, \& Gerhard 1995),
and 
outer stellar ring near its de Vaucouleurs radius 
(R$_{25}$ $\approx$ 330$''$).  
{\bf Figure 1} (extracted from the Digital Sky Survey using the SkyView 
utility [McGlynn, Scollick, \& White 1996])\footnote{NASA's SkyView facility
(http://skyview.gsfc.nasa.gov) was developed and is maintained under
NASA ADP Grant NAS5-32068 at NASA's Goddard Space Flight 
Center.}
shows the outermost portions of M94, highlighting the oval disk and outer 
pseudoring.
 
The inner starburst ring is a prominent 
source of H$\alpha$, 
H I, and CO emission (\markcite{smith91}Smith et al. 1991; 
\markcite{mulder93}Mulder \& van Driel 1993; \markcite{Gerin91}
Gerin, Casoli, \& Combes 1991).  
The discovery of compact thermal \& nonthermal 
radio sources in the ring (\markcite{Duric88}Duric \& Dittmar 1988) 
indicates the presence of 
dense H II regions and young SNRs.
The ring's velocity field can be described 
by 
a combination of circular rotation with velocities of 
order 200 km/s and residual non-circular motions 
of order 
15 km/s (\markcite{Mulder95}Mulder 1995) to 25 km/s 
(\markcite{buta88}Buta 1988), depending on the adopted 
inclination and major axis position angle.  

Interior to the ring, the bright bulge and inner disk 
show twisted isophotes at red and near-IR wavelengths, 
indicative of a weak bar-like distortion 
(\markcite{Beckman91}Beckman et 
al. 1991; \markcite{Shaw93}Shaw et al. 1993; \markcite{Mollenhoff95}
Mollenhoff et al. 1995).   
FIR and CO observations interior to the ring reveal a rich ISM 
with gas surface densities exceeding that of the ring 
(\markcite{Smith94}Smith \& Harvey 1994; 
\markcite{Garman86}Garman \& Young 1986; 
\markcite{Gerin91}Gerin et al. 1991; \markcite{Wong00}Wong \& Blitz 2000).

Optical spectroscopy of the nuclear region yields 
LINER-type emission lines along with absorption lines from the 
circumnuclear stellar population, consistent with an 
early main-sequence stellar turnoff (A4--A7) 
and corresponding age of $\sim$500 Myr (\markcite{Pritchett77}
Pritchett 1977; \markcite{Keel83}Keel 1983; 
\markcite{Taniguchi96}Taniguchi et al. 
1996).  
Further support for a 
young central population comes from NIR spectroscopy which shows deep CO 
absorption bands from red giant and asymptotic giant branch stars
of similar age (\markcite{Walker88}Walker et al. 1988).  These 
authors have proposed 
an outward propagating mode of star formation, whereby NGC 
253, M82, M94, and M31 represent increasingly evolved versions of 
the same starbursting sequence.
Although the kinematics of the ring show very little evidence for 
outward {\it expanding motions} 
(contrary to prior claims of bulk expansion 
[\markcite{kruit74}van der Kruit 1974; \markcite{kruit76}1976]), 
they also do not preclude a 
scenario for {\it radially propagating} star formation.  
Other investigators have modeled the inner and outer rings 
in terms of {\it resonant dynamics} mediated by bar or ``oval'' 
potentials interior to the rings 
(\markcite{Gerin91}Gerin et al. 1991; \markcite{Shaw93}
Shaw et al. 1993; \markcite{Mollenhoff95}Mollenhoff et al. 1995; 
\markcite{Mulder96}Mulder \& Combes 1996), with the observed
non-circular motions resulting from dispersion orbits near the 
Lindblad resonances (\markcite{buta88}Buta 1988).  

In this paper, we present and discuss new observational 
clues to the dynamical mechanisms governing the star formation
in M94.  
The ultraviolet images obtained by the Ultraviolet Imaging 
Telescope, in particular, 
reveal hitherto unrecognized 
patterns 
of recent star formation whose presence lends further support to the 
hypothesis of galaxy evolution via bar-mediated resonances.
The various imaging and spectroscopic observations and reductions 
are described in 
Section 2.  The resulting FUV, NUV, H$\alpha$, R, and I-band emission 
morphologies are 
presented and compared in Section 3.  Radial intensity profiles and other 
photometric comparisons are discussed in Section 4.  UV and optical 
spectroscopy of the inner 
disk and nucleus is presented in Section 5.  Kinematic properties and 
inferred dynamical scenarios are considered in Section 6.  Our summary of the 
wavelength-dependent morphological, spectro-photometric, and dynamical 
properties of M94
appears in Section 7, wherein 
evolutionary implications are discussed.

In the following Sections, we assume a distance to M94 of 4.6 (75/H$_{\circ}$) 
Mpc, based on the galaxy's recession velocity of 345 km/s 
with respect to the Local Group (\markcite{Sandage81}
Sandage \& Tammann 1981).  The corresponding 
spatial scale is 22.3 pc/arcsec.  Unresolved sources 
imaged by the HST/FOC provide additional constraints on the distance, as 
discussed in Section 4.  We adopt a nominal 
inclination of 40$^{\circ}$ and major-axis  
position angle of 120$^{\circ}$, while recognizing that both of 
these quantities may vary significantly with radius and with measuring 
technique (e.g. morphological vs. kinematic determinations) 
(\markcite{Bosma77}Bosma, van der Hulst, \& Sullivan 1977;
\markcite{buta88}Buta 1988; 
\markcite{Mulder93}Mulder \& van Driel 
1993; \markcite{Mulder95}Mulder 1995; \markcite{Mollenhoff95}Mollenhoff et 
al. 1995; \markcite{Wong00}Wong \& Blitz 2000). 

\section{Observations and Reductions}

A log of the ultraviolet and visible imaging is presented in {\bf Table 1}.
A listing of complementary UV and visible spectra is shown in {\bf Table 2}.

\subsection{Ultraviolet Imaging}

The Ultraviolet Imaging Telescope (UIT) imaged M94 in the far-ultraviolet
($\lambda_{\circ}$ = 1521 \AA, $\Delta\lambda$ = 354 \AA) on 1995 March 12
as part of the 16-day Spacelab/{\it Astro-2} mission aboard the Space Shuttle
Endeavour.  This wide-field telescope images 40-arcmin fields of view at 
$\sim$3 arcsec resolution.  In the case of M94, the UIT image represents 
the only extant UV image of the entire galaxy (see {\bf Figure 2a} 
(Plate xxx)).
The 1040-sec exposure was
obtained with a dual-stage image intensifier with CsI photocathodes
and was recorded on carbon-backed
IIaO Kodak film.  After processing of the film and scanning of the emulsion,
the
resulting digitized ``density image'' was fog-subtracted, flat-fielded,
linearized to ``exposure
units,'' and calibrated to flux units
using IUE observations of standard stars (cf. \markcite{Stecher92}Stecher
et al. 1992, \markcite{Stecher97}1997; 
\markcite{Waller95}Waller et al. 1995 and references therein).  
Correction for image
distortion produced by the magnetically focused image intensifiers
was carried out according to the procedures described by 
\markcite{Greason94}Greason et
al. (1994).  The resulting corrections amounted to a few arcsec in the field
center (which includes all of M94's FUV emission) 
increasing to 10--20 arcsec near the edge of the 40 arcmin field of 
view.  Astrometry was tied to 10 compact knots 
evident in both the FUV and B-band 
images (see next subsection).
Positions in the resulting distortion-corrected image are good to $\sim$3
arcsec, and the spatial resolution is of
similar magnitude.  

The Hubble Space Telescope's Faint Object Camera (FOC)\footnote{Data from 
the NASA/ESA {\it 
Hubble Space Telescope} were obtained at the Space Telescope Science Institute, 
which is operated by the Association of Universities for Research in Astronomy, 
Inc., under NASA contract No. NAS5-26555.} imaged the center of 
M94 in 
the near-ultraviolet ($\lambda_{\circ}$ = 2300 \AA, $\Delta\lambda$ = 500 \AA) 
on 18, July 1993 --- before the optical repair mission --- as part of a 
UV imaging survey of 110 large nearby galaxies 
(\markcite{Maoz95}Maoz et al. 1995, \markcite{Maoz96}1996).  After 
standard STScI pipeline processing, the 596 sec NUV exposure 
has a $22'' \times 22''$ field of view at 0.022$''$/pixel (see 
{\bf Figure 2a} (Plate xxx)).
The spherical 
aberration of HST's primary mirror resulted in a point-spread function (PSF) 
featuring a sharp core with FWHM $\approx$ 0.05$''$ and about 15\% of the total 
light surrounded by an extensive halo of 
several arcsec radius containing most of the energy.  
Following \markcite{Maoz96}Maoz et al. (1996), 
our calibration of the detected FOC counts into flux 
densities assume a conversion of $1.66 \times 10^{-17}$ erg s$^{-1}$ cm$^{-2}$ 
\AA$^{-1}$/count s$^{-1}$, while noting that the PHOTFLAM conversion in the 
image header is $2.017 \times 10^{-17}$ erg s$^{-1}$ cm$^{-2}$ \AA$^{-1}$/count 
s$^{-1}$.  Flux uncertainties are estimated at $\sim$5\% over 
large areas increasing to $\sim$20\% for compact sources 
(\markcite{Maoz95}Maoz 
et al. 1995, 1996, and references therein).

\subsection {Visible Imaging}

A wide-field ($9.66' \times 9.66'$) 
B-band image was obtained with the Palomar 
5-m telescope and Tek3 CCD camera (1024 $\times$ 1024 pix) 
on 1994 February 11 under hazy skies.  This 600-sec exposure is 
saturated in the central 2 arcmin, 
but contains high S/N detections of the oval 
disk and parts of the outer ring.  
Astrometry of this image is tied to the positions of several 
foreground stars, as 
measured on a corresponding image in The Digitized Sky 
Survey (\markcite{DSS94}STScI 1994)\footnote{The {\it Digitized Sky Survey} 
was 
produced at the 
Space Telescope Science Institute (STScI) under U. S. Government grant NAG 
W-2166.  The digitized images are based on photographic data 
from the Palomar Observatory Sky Survey (produced by 
the California Institute of Technology \& Palomar Observatory, and 
funded by the National 
Geographic Society)}.

Groundbased H$\alpha$, R, and I-band images of M94 were obtained with the 
now-retired KPNO 0.9-m telescope and RCA-1 CCD camera (508 $\times$ 316 pix) 
on 17, February 1986.\footnote{Kitt Peak National Observatory is operated
by the Association of Universities for Research in Astronomy, Inc., under
contract with the National Science Foundation.}  
These images have $7.28' \times 4.53'$ fields of view at 0.86$''$/pix.  Sky 
conditions varied from photometric to hazy, yielding PSFs of about 2$''$ 
(FWHM).  Astrometry is tied to 5 foreground stars that are common to the 
KPNO (H$\alpha$, R, I-band) and Palomar (B-band) images.
For calibration purposes, 
spectrophotometric standard stars (BD26$^{\circ}$2606 \& HD84937) 
were imaged before and after the target imaging.  Subtraction of the red 
continuum from the H$\alpha$-band image was carried out with the R-band image 
according to the formulations in \markcite{Waller9b}Waller (1990b), 
whereby corrections were made 
for the $\sim$38\% [N II]$\lambda\lambda$6548, 6584 contribution to the 
total H$\alpha$ $+$ [N II] line emission 
and resulting $\sim$15\% contamination of the 
H$\alpha$ image after transmission by the 36 \AA \ bandwidth H$\alpha$ filter.
A pure red-continuum image was also produced, by removing the contaminating
H$\alpha$ + [N II] line emission from the R-band image
\markcite{Waller90b}(Waller 1990b) 
(see {\bf Figure 2a} (Plate xxx)).  

Photometry of the H$\alpha$ 
emission from the $7.3' \times 4.5'$ field yields a total flux of 9.9 $\times$ 
10$^{-12}$ erg cm$^{-2}$ s$^{-1}$, 14\% higher than that determined by 
\markcite{Kennicutt83}Kennicutt \& Kent (1983) 
within a 7$'$-diameter aperture (after correcting for 
a 38\% [N II] contamination within their 100 \AA \ bandwidth).  
The starburst ring (R = 30$''$ 
$\rightarrow$ 60$''$) is measured to have f(H$\alpha$) = 9.6 $\times$ 
10$^{-12}$ erg cm$^{-2}$ s$^{-1}$, which is nearly 1.8 times higher than that 
obtained by \markcite{smith91}Smith et al. (1991) 
from the 75 \AA \ bandwidth image of \markcite{Pogge89}Pogge 
(1989).  Some of these discrepancies can be attributed to the varying 
bandpasses and corresponding uncertainties 
in the [N II] emission being transmitted,
vagaries in the continuum subtraction, 
problematic H$\alpha$ {\it 
absorption} produced by A-type stars in the inner disk (see Section 4.2),
and final calibration 
(\markcite{Smith91}Smith et al. 1991).
Because of hazy conditions during the I-band exposure, calibration 
of the I-band image was 
done by bootstrapping to the I-band 
radial intensity profiles resulting from the photoelectric photometry of 
\markcite{munoz89}Munoz-Tunon et al. (1989) and the 
$(R - I)$ color profiles of 
\markcite{Beckman91}Beckman et al. (1991).

\subsection{UV Spectroscopy}

FUV (1200 \AA \ to 2000 \AA) and NUV (2000 \AA \ to 3200 \AA) 
spectra of M94's inner disk were obtained from the International 
Ultraviolet Explorer (IUE) archive.
The IUE data were accessed via the IUE Data Analysis Center at 
NASA/GSFC\footnote{The IUE database is currently available at 
http://archive.stsci.edu/iue/} 
and 
are representative of the NEW Spectral Image Processing System (NEWSIPS).

{\bf Table 2} lists the image numbers, dates, exposure times, 
nominal positions 
and roll angles of the low-resolution FUV (SWP) and NUV (LWP \& LWR) IUE 
spectra.  Here, SWP, LWP, and LWR respectively refer to IUE's 
short-wavelength prime, long-wavelength prime, and long-wavelength redundant
cameras.
{\bf Figure 3} shows the IUE apertures ($20'' \times 10''$) on 
grey-scale images of the FUV emission, where the mapping is based on the 
nominal 
positions and 
roll angles of the FUV (SWP) and NUV (LWP \& LWR) observations.  
Although the accuracy of the nominal positions is of order 
$\pm$10$''$, {\bf Figure 3} indicates that the apertures were most likely 
sampling 
the inner disk rather than the starburst ring.  {\bf Table 2} notes which 
apertures are filled with disk (d) and/or nuclear (n) emission, based on 
visual inspection of the overlays.  The limiting 
spectral resolution of these data is about 6 \AA.

\subsection{Visible Spectroscopy}

Groundbased visible spectra (3800 \AA \ to 7500 \AA) were obtained from the 
Lick 
1-m Nickel telescope and Image Dissecting Spectrograph (IDS).  
The summed spectrum of the central region (8.1 arcsec 
circular aperture) is a mean-flux average of 64 min in blue and red grating 
settings, with 16 min in an intermediate setting to ensure the overlap area was 
well calibrated.  
The spectrum has been rebinned to 2.5-\AA \ pixels
from the original, which still oversamples the resolution of $\le$10 \AA \ 
FWHM.  We note that adjacent pixels in the IDS spectrum 
are not statistically independent, yielding detections similar to those of a
non-centroiding photon counter which spreads single photons across
several output pixels. 

We complement this composite 
spectrum with analysis of the image-dissector scanner
data presented by \markcite{Keel83}Keel (1983) and discuss 
a new high-resolution
spectrum, obtained with the 2.5-m Isaac Newton Telescope (INT) on
La Palma, using the Intermediate-Dispersion Spectrograph with image
photon-counting system (IPCS) detector.\footnote{The Isaac Newton Telescope is 
operated by the Royal Greenwich Observatory on behalf of the SERC at the 
Spanish Observatoriao del Roque de los Muchachos.}
 The 1" slit was oriented approximately
along the major axis (PA 135$^\circ$) for this 1000-second exposure.
There is useful signal over 45 spatial increments of 0.6$''$ each. We consider
equivalent widths of the lines along with spectral slopes and discontinuities
of the continuum emission.
 
\section{Ultraviolet and Visible Morphologies}

{\bf Figure 2a} 
(Plate xxxx) shows the dramatically different morphologies that 
are detected at 
ultraviolet and visible wavelengths.  The inner disk and bulge component 
that is so prominent at R-band completely disappears in the FUV image.  The 
oval disk at intermediate radius also has no FUV counterpart.  Instead, the 
FUV image is characterized by (a) an extended and elongated nucleus, 
(b) a 
diffuse inner disk, 
(c) a bright inner ring at the perimeter of the inner disk 
(R = 48$''$ = 1.1 kpc), and (d) two 500-pc size knots exterior to the ring (R 
= 130$''$ = 2.9 kpc). 

\subsection{Nuclear Region}

The insert in {\bf Figure 2a} (Plate xxxx) contains 
the HST/FOC pre-COSTAR image of the nuclear region 
(\markcite{Maoz95}Maoz et al. 1995, \markcite{Maoz96}1996).  
This near-UV image shows a marginally-resolved nucleus 
(FWHM $\approx$ 0.1$''$) 
embedded in a bright 
core of $\sim$1$''$ (22 
pc) diameter along with a faint 
``mini-bar'' that has a total length of 20$''$ (450 pc) 
at a position angle of approximately 30$^{\circ}$.  
Low-level ripples in the emission (which are included
in the estimate of the bar's total length) are evident 2$''$, 
5$''$ and 9$''$ to the 
SW of the nucleus.
Maoz et al. (1995) attribute these features to bow 
shocks or tidal arms resulting from a recent merger event.  However, we see 
that the ripples overlap with the larger concentric arcs 
that are evident in the UIT/FUV image at projected radii of 
$\sim$9$''$ and 15$''$ (see next subsection).  Similar ripples are also 
present in an archival HST/WFPC2 V-band image of this region -- with 
connections to larger spiral arcs.  
To the north of the nucleus are two point 
sources whose flux densities are consistent with those of single B-type 
supergiant stars (see Section 4).

\subsection{Inner Disk}

The UIT/FUV emission from the inner disk includes 
concentric FUV arcs to the SW of the nucleus which can be traced for 
$\sim$$\pi$ 
radians ($\sim$700 pc).  These arcs are of low contrast and, because
of the noise characteristics of the UIT imaging, are not amenable 
to typical contrast enhancement techniques (e.g. median filtering).  
However, they do show up as enhancements in the radial distribution of 
intensities (see {\bf Figure 2b}).
Typical spacing between the arcs is 9$''$ (200 pc), with 
additional features at 15$''$ and 34$''$.  The arcs themselves 
appear to show some substructure at the limits of resolution.

Contrary to a merging scenario (\markcite{Maoz95}Maoz et al. 1995), 
which would seem to be precluded by the lack of 
significant tidal effects beyond the inner few 
hundred parsecs, 
these features more likely 
indicate an orbital dynamic at work in the inner disk.  
At an interpolated orbital velocity of 
140 km/s and corresponding shear rate of -1360 km$^{-1}$ kpc$^{-2}$ 
(see Section 6), such features could have been differentially swept out over 
a timescale of 
$\sim$20--100 Myr.  This estimated timespan is consistent with a population
of late B-type stars whose main-sequence lifetimes are of similar duration.

The UIT/FUV image of 
the inner disk also shows 
a brighter arc to the north, just inside the starburst ring
at a radius of 40$''$ (see {\bf Figure 2b}), along with 
widespread diffuse emission at a level amounting to $\sim$15\% 
the mean surface brightness of the ring.  The nature of these resolved and 
unresolved FUV components is uncertain, although some clues can be obtained 
from 
the longer wavelength imagery.  

At H$\alpha$ the inner disk appears 
{\it in absorption} with respect to the continuum emission 
from the underlying 
population of stars.  
The strongest absorbers at H$\alpha$ are the photospheric atmospheres of 
B and A-type 
stars, whose temperatures are sufficiently cool for Hydrogen to 
remain neutral and sufficiently warm for significant population of the 
H Balmer (n=2) electronic energy level (\markcite{Mihalas78}cf. Mihalas 1978).
In the UIT/FUV image, the diffuse light from the inner 
disk probably arises from these same stars.  
Indeed, {\it the UIT/FUV image 
represents the first view of this young stellar population, unconfused by the 
longer-wavelength emission from the older inner disk and bulge components}.  

{\bf Figure 4} compares a spatially-filtered R-band image with an $(R - I)$ 
color 
image of the inner disk.  The spatially-filtered R-band image ({\bf Fig. 4a}) 
was created by 
median smoothing the R-band image over a $13'' \times 13''$ window and 
then subtracting the smoothed image from the original.  The resulting 
fine-scale structure includes 
the  
nuclear ``mini-bar'' previously noted by \markcite{mollenhoff95}
Mollenhoff et al. (1995) and \markcite{mulder95}Mulder 
(1995), along with a ``dark'' spiral arc of diminished 
emissivity to the west and other flocculent 
spiral structures 
associated with 
the starburst ring.  
The archival HST/WFPC2 V-band image of this region resolves the flocculent
structure into spiral dust lanes of high pitch angle which cross through the
nearly circular ring.

The $(R - I)$ color image ({\bf Fig. 4b}) 
shows reddening along the western 
spiral arc seen in ({\bf Fig. 4a}) and in another arc to the NE, 
along with blue knots in the 
starburst ring.  {\it None of the fine-scale structures and reddened features 
in the optical inner disk have counterparts in the UIT/FUV image.}  
Therefore, neither emitting nor absorbing structures, nor reddening dust 
lanes in the {\it old disk} can be defining the 
observed concentric FUV arcs to the SW and NE.  We infer that the FUV arcs 
are recently-generated structures of emitting B and A-type stars 
and/or scattering clouds of 
low dust optical depth.  We further speculate 
that the emitting and/or scattering
FUV sources, being relatively young, 
are distributed in a much thinner disk than that associated with 
the optical fine-scale structures and reddened features, 
thus explaining the morphological differences 
between these two wavelength regimes.
Further insights on the stellar content of the inner disk 
can be gained from the IUE spectroscopy presented in Section
6.

\subsection{Starburst Ring}
The inner starburst ring in M94 is the single dominant emission feature in the 
FUV, H$\alpha$, and radio continuum bands.  Rectification of the galaxy to 
its nominal face-on orientation (PA = 120$^{\circ}$ $i = 40^{\circ}$ [Mulder
\& van Driel 1993]) shows that the ring is almost
perfectly circular, with a mean radius of 47$''$ (1.1 kpc) and FWHM of 21$''$ 
(0.49 kpc).  The latter measurement is based on doubling 
the measured HWHM of the
emission {\it beyond} the mean radius, thereby excluding the diffuse
contribution interior to the ring. 
Adopting an inclination of 35$^{\circ}$, as in the dispersion
orbit model of Buta (1988), would yield a finite but small ellipticity
of 0.063, with a 16$^{\circ}$ offset between the kinematic line of nodes
and projected major axis.  We are not able to further constrain the ring's 
orientation, and so shall continue to assume an inclination of 
40$^{\circ}$, based on the H I study of Mulder \& van Driel (1993). 

A comparison of the FUV and H$\alpha$ morphologies 
reveals strong similarities.  {\bf Figure 5a} shows the H$\alpha$ emission, 
and {\bf Figure 5b} shows 
the ratio of 
H$\alpha$ and FUV intensities in the inner disk and ring.  Except for the 
spiral-like ridges of enhanced H$\alpha$ emission to the WNW and ESE, 
very little coherent 
variation in this ratio is evident in the ring.  In particular, the intensity 
ratios do not show any  
radial displacement in the FUV emission relative to that at H$\alpha$.
From this lack of radial structuring in the H$\alpha$/FUV ratio, 
we can conclude that 
{\it little evidence is found for outward or inward 
propagating starburst activity}, as 
explained below.
 
A morphological comparison of H$\alpha$ and FUV 
emitting regions provides a useful means of tracking 
sequential patterns of star formation in disk galaxies.  The H$\alpha$ 
emission is dominated by the most massive and short-lived stars (M $\ge$ 20 
M$_{\odot}$), while the FUV emission arises mostly from less massive, longer-
lived stars (20 $\le$ M/M$_{\odot}$ $\le$ 2.5).  For an evolving star cluster
with an IMF typical of local star-forming regions, the H$\alpha$ emission 
typically reaches a maximum within a million years of the star-forming 
episode, and quickly decays with an e-folding timescale of about 3 Myr.
By contrast, the FUV emission reaches a maximum at about 3 Myr 
(due to the onset of 
B-type supergiants), decaying to 1/e in 
yet another 3 Myr.  The H$\alpha$/FUV ratio is seen 
to decrease by 2 orders of magnitude
after 10 Myr (e.g. \markcite{hill95}Hill et al. 1995).  
Subsequently, the H$\alpha$ emission will
vanish, while the FUV declines slowly for the next few 100 Myrs.    

In the outward propagating starburst scenario, 
the H$\alpha$ emission would concentrate where the propagating 
wavefront is located --- on the outer perimeter of the FUV-emitting ring.
Inward propagating scenarios would have the H$\alpha$ emission interior to the
FUV ring.
Spatial 
displacements of H$\alpha$ and FUV emission have been observed 
across the spiral arms of M51 (\markcite{oconnell97}O'Connell 1997; 
\markcite{Petit96}Petit et al. 1996), 
M74 (\markcite{marcum97}Marcum et al. 1997), and NGC 4258 
(\markcite{Courtes93}Courtes et al. 1993).  These displacements have been 
interpreted as the result of spiral 
density waves concentrating gas along the insides of the spiral arms 
and subsequent migration 
of the evolving clusters past the spiral wave fronts.  Given the 
resolution of our images ($\sim$100 pc) and characteristic timescale between 
H$\alpha$ and FUV maxima (3 Myr), any residual 
propagation of starburst activity 
would have to proceed at a speed less than 35 km/s to 
 avoid detection.  Implications of this propagation speed limit are discussed 
in Section 7.  

The arm-like enhancements in the H$\alpha$/FUV ratio
correspond to similar features in a recent CO mapping with the BIMA 
interferometer (\markcite{Wong00}Wong et al. 2000).  
These arm-like extensions away from the ring
have further analogues at B-band, where a complex spiral pattern is evident
(see next subsection).

\subsection{Bi-Symmetric Knots and Associated Spiral Structure}
One of the most remarkable aspects of M94's FUV emission morphology is the 
pair 
of knots on diametrically opposite sides of the nucleus.  Both knots have 
projected radii of approximately 130$''$ (2.9 kpc).  A line drawn through the 
knots intersects the nucleus at a position angle of 105$^{\circ}$, 
intermediate between the PAs of the inner starburst ring (127$^{\circ}$) and 
the oval disk (95$^{\circ}$) (\markcite{mollenhoff95}Mollenhoff et 
al. 1995).  The two knots are roughly 20$''$ (500 pc) in size, 
with the eastern knot 
showing the more complex structure.  Only the eastern knot shows significantly 
in our 
continuum-subtracted H$\alpha$ image ({\bf Figure 5}).  
Other much smaller knots of FUV emission 
appear at low S/N to the SW, SSW, ENE, and NNE.  The SW, SSW, and ENE knots 
have faint counterparts in the light of H$\alpha$, while all of these 
features have 
counterparts at B-band (see below).

{\bf Figure 6a} and {\bf Figure 6b} respectively show 
the B-band image before and after spatial 
filtering.
The filtering in this case involves median smoothing 
over a $30'' \times 30''$ window and then dividing the original image by this 
smoothed image.  Such {\it median normalized spatial filtering} 
reveals fine-scale structure over a wide range of surface brightnesses (e.g. 
\markcite{Waller98}Waller et al. 1998).  
Here, it highlights the starburst ring and bisymmetric 
knots as significant enhancements above the disk at B-band.  
The filtered image 
also reveals many flocculent 
spiral arms outside of the ring, whose relation with the 
knots is somewhat ambiguous.  The western knot appears part of a dominant arm 
with a ``shingled'' morphology connecting to the ring in the northeast.  
The eastern knot appears to be associated with several arms, including one 
which would be the symmetric counterpart to the arm that links the western 
knot 
to the ring.  The marginally detected FUV knots to the SW, SSW, ENE, and NNE 
all show enhanced B-band emission from associations of massive 
stars 
among the myriad spiral arms.  

In summary, the bi-symmetric FUV knots have B-band counterparts 
which appear to be part of 
the complex spiral arm structure at these radii.  Their 
symmetric prominence on opposite sides of the nucleus, however, requires a 
dynamical explanation that is spatially more specific 
than that of the arms.  Similarly, the 
starburst ring can be regarded as an especially bright, tightly wound 
component of the overall spiral pattern, whose prominence also indicates 
special dynamical circumstances.

Bi-symmetric knots or plumes have been noted in other ringed-barred 
galaxies, including NGC 1433 (\markcite{buta86}Buta 1986), 
NGC 7020 (\markcite{buta90}Buta 1990), 
NGC 7098 (\markcite{buta95a}Buta 1995), and
IC 4214 (\markcite{bc96}Buta \& Combes 1996).  As discussed in 
these papers, the symmetric features are likely tracing dynamical resonances
that are connected with the general ring-bar phenomenon.

\section{Photometric Results}

\subsection{Nuclear Photometry}

From the UIT and HST/FOC images, 
circular aperture photometry of the nucleus out to a radius of 
5$''$ yields $m(FUV)$ = 14.7 mag and 
$m(NUV)$ = 13.5 mag, respectively.  
The resulting $(FUV - NUV)$ 
color of 1.2 mag is significantly redder than those 
derived from IUE spectra of the inner disk (see next Section).  Between 
5$''$ -- 10$''$ radii, the non-nuclear emission is significantly bluer 
than those 
derived using IUE, with 
$(FUV - NUV) \approx -0.3$.  The latter 
colors indicate the presence of late B and 
hotter stars, depending on the amount of reddening 
(\markcite{Fanelli92}Fanelli et al. 1992).  Estimates of the reddening 
have been derived from the visible colors (\markcite{Smith91}Smith et al. 1991;
see also next subsection), visible spectra and 
subsequent modelling of the stellar populations (\markcite{Taniguchi96}
Taniguchi et al. 1996; 
\markcite{Pritchett77}Pritchett 1977).  The resulting estimates of 
visual extinction are 0.5--1.0 mag in the nuclear region, 
corresponding to UV color excesses of $E(FUV - NUV)$ = 0.1--0.2 mag.
Therefore, the corrected $(FUV - NUV)_{\circ}$ 
color of the 
nucleus could be close to 1.0 mag, or the equivalent of a 
stellar population with a 
late A-type stellar cutoff (\markcite{Fanelli92}Fanelli et al. 1992).
 
The non-nuclear point sources in the HST/FOC near-UV 
image of the nuclear region provide 
helpful checks on the distance to M94.
For the brightest point source, \markcite{Maoz96}Maoz et al. (1996) 
lists a monochromatic flux 
of $f_{\lambda}$(2300 \AA) = 2.6 $\times$ 10$^{-16}$ erg s$^{-1}$ cm$^{-2}$ 
\AA$^{-1}$ (m$_{\lambda}$(2300\AA) = 17.9 mag).  
This source is unresolved at a 
resolution of 0.1$''$ and corresponding linear scale of $<$2.2 pc 
at the assumed 
distance.  Although an extremely compact star cluster 
cannot be ruled out, this 
exceptional UV source is most likely dominated by a single hot supergiant star.
In the absence of extinction, a B0-2 Ia-O 
supergiant star has $M(NUV)$ $\approx$ $-$10.7 mag 
(\markcite{fanelli92}Fanelli et al. 
1992), 
which 
would imply a distance modulus of ($m - M$) = 28.6 mag 
and corresponding distance 
of 5.2 Mpc.  
Assuming nuclear color excesses of $E(B-V)$ = 0.15--0.27 mag 
(\markcite{taniguchi96}
Taniguchi et al. 1996) and Galactic-type 
extinction law, the corresponding near-UV extinctions would be 
$A(NUV)$ $\approx$ 1--2 mag, and the revised distances would be 
3.3--2.1 Mpc, respectively.  The adopted 
distance of 4.6 Mpc can thus be regarded as a reasonable estimate, given the 
uncertainties in stellar spectral type and extinction.      
The second source, 1-arcsec to the E, is 6 times (1.9 mags) fainter -- 
consistent with it 
being an A0-2 Ia-O supergiant at this distance, 
if subject to the same amount of extinction.  Were the galaxy significantly 
closer, more abundant OBA-type 
giant and main-sequence stars would be resolved -- 
which is not the case.

\subsection{Surface Photometry of the Disk}

{\bf Figure 7} shows radial profiles of the FUV surface brightness and 
cumulative flux.  These profiles were derived from 
annular-averaged photometry using elliptical annuli consistent with the 
adopted position angle (120$^{\circ}$) and inclination (40$^{\circ}$).  
Since Galactic extinction is 
negligible towards M94 (\markcite{rc3}de Vaucouleurs et al. 1991), 
no correction was made.  
The resulting surface brightness profile ({\bf 
Fig. 7a}) is dominated by light from the nucleus, inner disk, starburst ring, 
and an exponentially-declining disk that is punctuated by an enhancement from 
the 
bi-symmetric knots at $R = 130''$.  
The exponentially declining component between 50 and 80-arcsec radii 
has an e-folding scalelength of 
only $10''$ (223 pc), similar to that found at R and I-bands (cf. 
\markcite{munoz89}Munoz-Tunon 
et al. 1989).  The FUV emission beyond the bi-symmetric knots also shows a 
steep decline with an estimated scalelength of $20''$ (446 pc), significantly 
shorter than the 70$''$ scalelength measured at R and I-bands in the 
outer disk 
(Munoz-Tunon et al. 1989).

The cumulative flux profile ({\bf Fig. 
7b}) indicates that the half-light radius matches that of the starburst 
ring, and that more than 80\% of the total FUV emission is contained 
within a radius of $60''$ (1.3 kpc).  The total FUV flux from M94 
is f$_{\lambda}$(1520 \AA) = 
6.02 $\times$ 10$^{-13}$ erg cm$^{-2}$ s$^{-1}$ \AA$^{-1}$, 
corresponding to m(FUV) = 9.45 mag, or M(FUV) = $-18.91$ mag at the adopted 
distance of 4.6 Mpc.  

The corresponding FUV luminosity of 1.5 $\times$ 10$^{39}$ 
erg s$^{-1}$ \AA$^{-1}$ 
is the photometric equivalent of 2 $\times$ 10$^4$ Orion nebulae 
(\markcite{Bohlin82}Bohlin et al. 1982) or 
about 90 30-Doradus regions (as measured on an UIT/FUV image of 30 Dor 
out to a radius of 
5 arcmin [67 pc]), before correcting for the extinction in these sources.  
Adopting a 
Salpeter IMF with lower and upper mass limits of 0.1 and 100 M$_{\odot}$, 
respectively, yields an uncorrected star formation rate of 0.15 M$_{\odot}$/yr.
This SFR estimate assumes continuous star formation and 
includes the strong contribution of B supergiant stars to the overall 
luminosity. 
An overall visual extinction of 1 mag would increase the global luminosity and 
inferred SFR by about a factor of 11 (\markcite{Hill97}Hill et al. 1997).  
At this rate, it would have taken $\sim$10 Gyrs to transform 
the dynamical mass of 
1.6 $\times$ 
10$^{10}$ M$_{\odot}$ that is present 
within $60''$ of the nucleus(\markcite{Garman86}Garman \& Young 1986) 
into the dominant stellar component that we see 
today.  

As a check on these starbirth estimates, we note that the measured H$\alpha$ 
flux of 9.6 $\times$ 10$^{-12}$ erg cm$^{-2}$ s$^{-1}$ converts to a 
luminosity 
of L(H$\alpha$) 
= 2.45 $\times$ 10$^{40}$ erg/s, or only 5600 equivalent Orion nebulae.  The 
origin of this discrepancy is deferred to the next subsection.
Assuming Case B recombination and multiplying the H$\alpha$ luminosity by 
$7.4 \times 10^{11}$ yields a photoionization rate of N$_i$ = 1.8 $\times$ 
10$^{52}$ photons/sec, and corresponding star formation rate of 0.22 
M$_{\odot}$/yr.  The similarity of the FUV and H$\alpha$-based SFRs, 
{\it before correcting for 
extinction}, suggests either that insignificant obscuration is present in the 
photometrically dominant starburst ring, 
or that significant obscuration 
exists with other mechanisms making up for the greater attenuation of the FUV 
emission relative to the longer-wavelength H$\alpha$ emission.  
Such mechanisms include (1) an excess  
contribution of non-ionizing B-type stars to the total FUV emission, 
(2) a reduction of the total H$\alpha$ emission due to H$\alpha$ absorption by
the atmospheres of B and A-type stars, and (3) 
the absorption of EUV photons before they ionize the gas and 
induce H$\alpha$ emission.  Likely EUV absorbers include the nebular dust 
associated with the H II regions as well as the 
various metal species in the O-type stellar 
atmospheres themselves (cf. \markcite{Hill97}Hill et al. 1997; 
\markcite{Waller96}Waller, Parker, \& Malumuth 1996).  
We conclude in the next subsection that 
modest obscuration plus non-ionizing contributions to the total FUV emission
best explain the observed levels of H$\alpha$ and FUV emission.
Stellar absorption at H$\alpha$ is probably less than a 10\% effect overall,
given H$\alpha$ emission 
equivalent widths of order 100 \AA \ in the ring, and stellar 
absorption equivalent widths peaking at less than 10 \AA. 

The global (total) $(FUV - V)$ color of M94 is 1.21 mag, which is 
characteristic of early-type 
disk galaxies with circumnuclear starburst activity (e.g. NGC 1068, 
NGC 3351 [\markcite{Waller97}Waller et al. 1997]).
The $(R - I)$ colors shown in {\bf Figure 4b} do not vary as much as those 
found 
by Beckman et al. (1991) in their photometric study.  Bootstrapping our 
measured intensity ratio in the nuclear region to a color of $(R - I)$ = 
0.45 mag 
as 
reported by Beckman et al. (1991), we obtain colors that range from $(R - I)$ 
= 0.3 mag in the rings' starburst knots to $(R - I)$ = 0.5 mag 
in the arcs interior to the ring.
Beckman et al. (1991) 
obtain much bluer colors of order 0.1 mag near the ring.  We 
attribute this discrepancy to our having used the H$\alpha$-band image to 
remove contaminating line emission from the R-band image (\markcite{Waller90b}
Waller 1990b), thereby reducing the red flux from the knots in the ring by 
20\%.  Comparison of the contaminated and uncontaminated R-band images 
confirm our attribution.

If 
the reddened arcs are caused by dust, the corresponding color excess relative
to that of the starburst knots would 
amount to $E(B - V) \approx E(R - I) = 0.2$ mag, 
or the equivalent visual extinction of roughly 0.6 mag in excess of the knots. 
Allowing for bluer stellar populations in the knots could reduce the estimate 
of excess 
extinction 
in the arcs to negligible levels, 
while inclusion of some internal reddening in the knots would increase the 
total estimated extinction by $\le$1 mag.
Such low estimates for the extinction 
are similar to the spectroscopic results obtained by Pritchett (1977) and by 
Taniguchi et 
al. (1996) in their studies of the nuclear region, 
where color excesses of 0.27 
mag and 0.15 mag, corresponding to $A_V = 0.86$ mag 
and 0.48 mag, were respectively obtained.

\subsection{Photometry of the Starburst Ring and Knots}

Photometry of the FUV and H$\alpha$ emission from the starburst ring 
(between projected radii of 30$''$ and 60$''$) yields a mean 
H$\alpha$/FUV flux ratio (in equivalent width units) of 22 \AA.  
By way of comparison, we note that this flux ratio 
is somewhat lower than that found in the Orion nebula 
(38 \AA [Bohlin et al. 1982]), within the range of 
flux ratios evident in M33's 
giant H II regions (19 \AA \ to 69 \AA \   
[\markcite{Parker96}Parker, Waller, \& Malumuth 1996]), 
and close to the mean of M51's 
wide-ranging GHRs ($29 \pm 23$ \AA \ [\markcite{Hill97}Hill et al. 1997]).

A model of 
steady-state star formation with a Salpeter-type IMF  
yields a somewhat lower ratio of 16 \AA, 
while a modeled 2 Myr-old starburst would match our result
(\markcite{Hill95}Hill et al. 1995 [see their Fig. 10]).  
Such a short burst timescale is probably 
untenable, thus indicating some need to correct our FUV flux for extinction. 
For 1 mag of visual extinction, the resulting ``corrected'' 
H$\alpha$/FUV flux ratio would be lower by a factor of $\sim$5, leading to 
a modeled burst age of 6 Myr --- close to 
the age limit of the ionizing O stars.  
Based on these considerations alone, we surmise 
that the FUV and H$\alpha$
emission from the starburst ring are subject to no more than 1 mag of visual 
extinction (on average).  We also can conclude that the similarity in 
uncorrected SFRs based on the global FUV and H$\alpha$ fluxes is probably 
coincidental.  After correction for 0.5--1.0 mag of visual extinction, 
the H$\alpha$ 
flux would be increased by a factor of 1.4--2.1, while the FUV emission would 
increase by a factor of 3.3--11.  In other words, 
{\it the FUV emission is tracing 
an additional non-ionizing Population I component in the inner disk that is 
missed at H$\alpha$ and, after correction for extinction, 
is contributing to the overall calculated 
rate of star formation}.

The bi-symmetric FUV knots to the west and east of the ring 
have fluxes of $3.0 \times 10^{-15}$ 
and $3.8 \times 10^{-15}$ erg cm$^{-2}$ s$^{-1}$ \AA$^{-1}$, respectively.  
This translates to the equivalent of 100 and 127 Orion nebulae, respectively.  
Only the 
eastern knot shows significant H$\alpha$ emission, 
with a flux of 2.57 $\times$ 
10$^{-13}$ erg cm$^{-2}$ s$^{-1}$ and hence an H$\alpha$/FUV flux ratio of 67 
\AA --- about twice that of the Orion nebula --- thus indicating greater FUV 
obscuration than is present in Orion or a hotter ionizing cluster.  
Examination 
of the spatially-filtered B-band image ({\bf Figure 6b}) reveals 
dark
spiral-arm features at the position of the eastern knot which may 
correspond to 
obscuring dust clouds.  The western knot shows similar structures to one side.
The lack of significant H$\alpha$ emission from this knot is difficult to 
explain with dust, and is most likely due to the presence of an older ($\tau > 
10$ Myr), 
non-ionizing stellar population.  

\section{Spectroscopic Results}

\subsection{FUV Spectroscopy}

{\bf Figure 8a} shows an average of 5 IUE/SWP spectra of the extranuclear 
emission from the inner disk.  Based on the aperture mapping shown in {\bf 
Figure 3a} some 
emission from the nucleus may be present, but at low levels.  The wavelengths 
of commonly observed 
FUV absorption and emission lines are indicated on the spectrum for comparison.
Although the averaged spectrum is of generally low S/N, 
several features can be 
identified.  Of these, the most prominent are the absorption blends of 
S II (1250-1259) and Si II 
(1260,1265), the P-Cygni profile of S IV 
(1394,1443), the absorption blends of 
Fe III (1601-1611), Al III (1600-1612), 
C II (1720-1722) and Al II (1719-1725), and part of 
the broad absorption complex of Fe III (1891-1988).  The 
low-ionization lines, in particular, are most characteristic of 
B and A-type stars. The strength of the blueshifted 
S IV absorption feature relative to 
that of C IV (1550) along with the strong Si II absorptions at 
1260,1265 \AA \ and
1527,1533 \AA \ 
indicate a composite spectral type later than B3 but earlier than B8 
(Fanelli et al. 1992; Kinney et al. 1993; Walborn, Parker, \& Nichols 1995). 
The absorptions at 1470 \AA, 1780 \AA, and 1790 \AA \ remain unidentified.
  
In emission, there is some evidence for C IV (1548-1551), 
He II (1640), N III] (1730,1750), and C III] (1909) --- much of which may be 
the result of LINER activity being scattered by the ISM in the inner disk.
The C III] (1909) line emission may be responsible for 
filling in part of the broad stellar 
absorption complex of Fe III (1891-1988).

\subsection{NUV Spectroscopy}
{\bf Figure 8b} shows an average of 2 IUE/LWP spectra of the disk.  
The most prominent
spectral features are the 
absorption lines of Fe II (2609, 2750), Mg II (2800), and Mg I (2852). 
Comparison with stellar spectra
from the IUE Spectral Atlas (\markcite{wu91}Wu et al. 1991) 
shows that Mg II (2800) is
unusually weak relative to Mg I (2852); the line ratios in this
spectral range being consistent  
with light dominated by late G-type stars, based on this cursory 
comparison.  However, the
spectrum is too blue and the amplitude of the 2800 \AA
\ break too small for this to be the whole story. Comparison with the
nuclear spectra (IUE LWR 12221,12238 -- as listed in Table 2)
shows evidence for a spectrally smooth blue component off the nucleus
proper. Also, the Mg II 2800 \AA \ break is roughly 50\% smaller than that
evident in the 
nuclear spectrum (see below).

All the absorption features are broader in
the off-nuclear spectrum, because the light there almost uniformly fills the
aperture (so centering shifts are not an issue). The off-nuclear spectrum 
shown here is
not only bluer but shows a diminished 2800 \AA \ break amplitude. Measured as a
ratio of flux F$_{\lambda}$ above and below 2800 \AA, 
the ratio is 2.5 at nucleus and
1.8 away from it. Using the spectral-break indices developed by Fanelli et al.
(1992), we obtain for the off-nucleus averaged spectrum 
(Mg II 2800 -- Mg I 2852) = 0.31 mag, (2609/2660) = 0.45 mag,
and (2828/2921) = 0.31 mag --- all of which are consistent with a $(B - V)$ 
color of 0.2--0.3 mag, or the equivalent of a late A-type main sequence star 
or late A to early F-type giant/supergiant star.

The off-nuclear continuum is likewise flatter in the SWP
range, 
 with little trace of the emission-like features around 1900 \AA \ 
seen in the nuclear
spectrum (see \markcite{Kinney93}Kinney et al. 1993). 
These properties can be
accounted for, if recent star formation
(few times $10^8$ years ago) contributes relatively more light off the nucleus
than on it, consistent with an aging burst that is spatially more extended 
than the
centrally-concentrated stars of the older disk and bulge.

\subsection{Optical Spectroscopy}
{\bf Figure 8c} shows the spectrum of optical emission from the central 
D = 8.1 arcsec, as obtained with the Lick 1-m telescope and IDS.
Absorption lines include the Ca II K\&H lines at 3934 \AA \ and 3968 \AA, 
H$\delta$ (4101), CH G-band (4300), H$\gamma$ (4340), He II (4686), H$\beta$ 
(4860), Mg I (5170), Fe I (5270), Na D (5890,5896), H$\alpha$ (6563), and 
atmospheric absorption bands of O$_2$ (6867) and H$_2$O (7186).
Emission lines are 
restricted to [N II] (6584) and [S II] (6731).

The H$\beta$ equivalent width of
about 5.4 \AA \ is huge for an old population typical of spiral bulges, 
even before any correction for
line emission. There has been star formation in this region not long ago. 
The 4000-\AA \ break is also suppressed, indicating the effective age is
much younger than for typical early-spiral bulges. Following the
\markcite{dressler87}
Dressler and Shectman (1987) definition (ratio of F$_{\nu}$ between
3950-4050 and 3750-3850 \AA), the observed break has F$_{\nu}$ = 1.6, 
whereas values of 1.9--2.0 are more usual.

From the IDS/IPCS on the 2.5-m Isaac Newton Telescope, we see that
the nucleus (inner 1" or so) is bluer at optical wavelengths 
than its immediate surroundings.
Misalignment in the IPCS detector is less than 0.1 pixel, 
so this is not an obvious
instrumental effect. This nuclear 
blueing effect is substantial, amounting to
about 30\% over the 4400--5200 \AA \ range 
(more or less equivalent to $E(B-V) =
-0.3$ mags, where the mean ($B - V$) color of the bulge is 0.91 mags, which is 
already bluer than ordinary Sa/Sb bulges \markcite{keel78}[Keel \& 
Weedman 1978]).
Outside this area, the color along the slit is quite constant out to
15$''$ from the nucleus, where the signal begins to die out.  This effect has
no observed counterpart in the IUE spectra, because of their courser spatial
resolution.

The prominent H$\beta$ absorption line in the nuclear INT/IPCS spectrum 
has FWHM 18--20 \AA, consistent with
values seen in mid-to-late A stars. Thus, the data are consistent with
seeing the main-sequence turnoff near this spectral type, rather than
the supergiant dominance of a much younger population giving superficially
similar spectral features (as seen in the starbursting nucleus of 
NGC 4569 [\markcite{Keel96}Keel 1996]).

The optical spectra confirm our imaging result that H$\alpha$ is 
in absorption with respect to the continuum from  
the underlying population of A and B-type stars.  By contrast, the neighboring 
line of [N II] (6584) is seen in emission in the Lick/IDS spectrum.  
As previously noted, 
H$\beta$ is also dominated by 
photospheric absorption.
The IDS/IPCS spectrum shows some H$\gamma$ emission at least at the nucleus, 
that is not swallowed by the stellar absorption line. 
 
\section{Kinematics and Dynamics}

{\bf Figure 9a} shows the rotation curve based on the H I observations of 
Mulder 
and van Driel (1993)\markcite{Mulder93}, where key morphological features are 
noted.  This curve is qualitatively similar to that obtained from a recent
interferometric mapping of CO (\markcite{Wong00}Wong \& Blitz 2000) but is 
on-average 12\% lower, due to the adoption of a 40$^{\circ}$ inclination
compared to their 35$^{\circ}$.  {\bf Figure 9b} 
shows the corresponding radial profiles of H I and H$_2$ 
gas, the latter being derived from the CO observations of Gerin et al. 
(1991)\markcite{Gerin91}.  

The molecular gas component 
clearly dominates the inner disk's 
ISM and may continue to prevail at higher radii, where the CO emission 
has yet to be 
measured.  The plotted extrapolation beyond the last reliable measurement of 
CO emission
is intended to provide an {\it upper limit} on the total gas and thus a lower
limit on the Q stability index (see below).  The interferometric measurements
by Wong \& Blitz (2000) yield even lower extrapolated surface densities of 
gas -- consistent with our extrapolated upper limit.  Moreover, FIR 
measurements with IRAS and the KAO (\markcite{Smith91}Smith et al. 1991) 
reveal negligible dust emission beyond 60 arcsec radius -- further 
corroborating the gas upper limit used here.  
The H I profile, though of lower amplitude, 
shows enhancements at the
radii of the starburst ring and the bi-symmetric knots.  

{\bf Figure 9c} shows the corresponding radial profile of the 
gravitational stability index (Q).  Here, we have considered the simplest case 
of the gaseous stability, without any coupling with the stellar component, such 
that

$$Q = \Sigma_{crit}/\Sigma_{gas} = \kappa \sigma/ \pi G \Sigma_{gas},$$

\noindent
where the epicyclic frequency $\kappa$ 
is closely linked to the rotation curve v(R) 
through

$$\kappa^2 = {{2v}\over{R}}({{v}\over{R}} + {{dv}\over{dR}}),$$

\noindent
and where $\sigma$ is the gas velocity dispersion --- a quantity which is not 
well constrained but is probably of order 5--10 km/s.  Here, we leave $\sigma$ 
as an unknown.
The resulting radial profile of Q/$\sigma$ 
out to the limits of the CO observations shows a 
shallow minimum at the radius of the starburst ring.
Although this minimum is too broad and shallow 
to explain the more discrete and prominent starburst ring, it may 
help to explain the relatively young ($10^7 - 10^8$ yr) stellar population 
pervading the inner disk.

{\bf Figure 10} shows the orbital resonance diagram that results from the 
rotation curve along with the radii of key morphological features.
A single 
wave pattern speed of 35 km s$^{-1}$ kpc$^{-1}$ (where d = 4.6 Mpc) 
would place the nuclear bar 
inside the Inner-Inner Lindblad Resonance (IILR), 
the starburst ring between the IILR \& Outer Inner Lindblad Resonance (OILR), 
the bisymmetric 
knots at the 4:1 ``ultra-harmonic'' resonance, 
the oval disk terminating at co-rotation (CR), 
and the outer pseudo-ring at the 
Outer Lindblad Resonance (OLR).  
Alternatively, a higher pattern speed of 56 km s$^{-1}$ kpc$^{-1}$ (as 
modeled by \markcite{mulder96}
Mulder \& Combes (1996)) would move the starburst ring just outside 
the OILR, the bisymmetric knots to the CR radius, 
and the edge of the oval disk 
to the OLR.  This latter model, however, fails to account for the outer
ring without invoking additional patterns.

Further support for the proposed sequence of resonances comes from specific
ratios of the corresponding radii (cf. \markcite{buta86}Buta 1986).
Given a flat rotation curve, the modeled ratio of OLR and UHR radii is 
r(OLR)/r(UHR) = 2.6.  If these OLR and UHR resonances are respectively 
traced by the outer pseudoring and bi-symmetric knots, then we obtain a ratio
of radii equaling 2.54 -- closely matching the modeled ratio.  The theoretical
ratio of OLR and CR radii, being r(OLR)/r(CR) = 1.7, is also well matched
by the observed relative dimensions of the outer ring and oval disk, 
where a ratio of 1.5 is obtained.

Admittedly, morphological tracers and rotation curves are 
insufficient to discriminate between these and other possible kinematic 
patterns (cf. \markcite{bc96}Buta \& Combes 1996). 
Further progress on constraining the resonant dynamics in M94 will 
require analysis of the complete velocity field in the disk (cf. 
\markcite{westpfahl96}Westpfahl \& 
Adler 1996; \markcite{canzian97}Canzian \& Allen 1997).  
A 
complete H I mapping with the VLA has been made recently, 
whose spectral and spatial 
resolution would be sufficient to derive a complete velocity field and its 
associated resonant states
(D. Westpfahl, private communication).   
Until such an analysis is carried out, we think that the resonant state
diagrammed in {\bf Figure 10} best explains the most features with the
fewest conditions.

\section{Conclusions and Implications}

Through UV-Optical imaging and spectroscopy, we have found new evidence for 
bar-mediated resonances as the primary drivers of evolution in M94 at the 
present epoch.   Our observational results include evidence for 

\noindent\hang(1.)  A 450-pc long nuclear ``mini-bar'' at both 
optical and near-UV wavelengths.  

\noindent\hang(2.)  An inner disk with diffuse FUV emission in concentric arcs 
that do not match the fine-scale structures or reddened structures at visible 
wavelengths.  Since $H\alpha$ is observed in absorption here, 
the UIT/FUV image 
represents the first view of this non-ionizing but relatively young 
disk component.

\noindent\hang(3.)  UV-Optical colors and spectral indices 
in the nucleus and inner disk that indicate
B and A-type stars in the presence of modest extinction 
(A$_V$ $\le$1 mag) along with some LINER activity from the nucleus itself.

\noindent\hang(4.)  A 2.2 kpc diameter 
starbursting ring at the perimeter of the inner disk that 
is bright at FUV, H$\alpha$, and radio-continuum wavelengths.
The level of starbirth activity in this inner ring rivals the levels 
observed in 
starbursting irregular galaxies such as NGC 1569 and NGC 4449.  The inferred 
star formation rate within the ring and inner disk amounts to 1.5 
M$_{\odot}$ yr$^{-1}$ --- sufficient to build up the stellar mass of the 
inner disk and bulge in $\sim$10$^{10}$ yr.
  
\noindent\hang(5.)  No detectable 
radial offsets between the H$\alpha$ and FUV rings, thus 
indicating a 35 km/s speed limit to any outward or inward propagating star 
formation in the ring, if such a mode is present.

\noindent\hang (6.) Two 500-pc size FUV-emitting knots exterior to the ring on 
diametrically opposite sides of the nucleus.
The bisymmetric knots and starburst ring appear to be especially prominent
parts of a complex spiral arm structure, as revealed in a spatially-filtered
B-band image.

\noindent\hang (7.) The starburst ring, 
bi-symmetric knots, oval disk, and outer pseudo-ring as signposts of resonant
dynamics in the disk of M94.  More specifically, the radii of these features 
match those of various orbital resonances, given a pattern speed of 35 km 
s$^{-1}$ kpc$^{-1}$ at our adopted distance and inclination.
These orbital resonances 
are most likely driven by some combination of the nuclear mini-bar and 
oval distortion in 
the disk.  

\noindent\hang (8.) A shallow minimum of gravitational stability at the radius 
of the starburst ring that extends inward into the inner disk.  
Although too broad 
to explain the discrete 
starburst ring, the shallow minimum may help to explain the 
10$^7-10^8$-yr old stellar population interior to the ring.  
\vskip 12pt

Although we can set a limit on the speed of outward or inward 
propagating star formation 
in the ring, we cannot preclude the existence of such a mode.  At a 
propagation 
speed of 35 km/s, a wave initiated in the nucleus 
could traverse the inner disk 
to the radius of the current starburst ring in only 31 Myr.  Therefore, it is 
possible that the $10^7-10^8$-yr old stellar population detected in the inner 
disk is the result of such an outward propagating wave.  The striking 
difference in emission morphologies at FUV and 
red wavelengths provides further 
support for the starburst ring being a transient phenomenon which does not 
persist at any one radius for very long.  Either these resonant phenomena come 
and go, as the oval distortions undergo secular evolution, or their operating 
radii migrate in response to other dynamical influences on the stars and gas 
(cf. \markcite{Combes94}Combes 1994; \markcite{Combes95}Combes et al. 1995; 
\markcite{Friedli95}Friedli \& Benz 1995).  Otherwise, one must invoke strong
radial inflows of stars from the starburst ring to populate 
the inner disk and bulge,
a feat requiring unusual circumstances -- e.g. mergers.

The results reported herein may have important 
implications with regard to observations of the most distant 
observable galaxies.  At redshifts of 1--5, the 2-kpc diameter 
starburst ring in M94 would subtend 
angles of only (0.7$''$ -- 1.0$''$)H$_{\circ}$/75 in an Einstein-de Sitter
Universe (q$_{\circ}$ = 1/2) and (0.3$''$ -- 0.2$''$)H$_{\circ}$/75 in an
open (Milne) universe (q$_{\circ}$ = 0) (cf. 
\markcite{Narlikar83}Narlikar 1983).  
The UV-bright nuclear rings evident in NGC 1097, NGC 1317, NGC 1433, 
NGC 1512, NGC 2997,  
NGC 4321, and NGC 5248 
(\markcite{maoz95}Maoz et al. 1995; \markcite{maoz96}Maoz et al. 
1996; \markcite{kuchinsky00}Kuchinsky et al. 2000; 
\markcite{marcum00}Marcum et al. 2000) would subtend even smaller angles at
the same redshifts.  Moreover, nuclear rings tend to have higher FUV surface
brightnesses  
than their larger counterparts -- the inner ring in M94 being a remarkable
exception.
Therefore, some
of the ``core-halo'' morphologies that are evident at high-redshift in the 
restframe FUV (cf. \markcite{Giavalisco97}Giavalisco et al. 1997) may, 
in fact be marginally-resolved 
representations of galaxies with starburst rings in their centers.  

Gravitationally-lensed galaxies are fortuitously magnified, enabling 
resolutions of their structure at high S/N.  An important precedent in this 
regard is the gravitationally-lensed ``Pretzel Galaxy'' which lies behind the
galaxy cluster 0024+1654 at an estimated redshift of 1.2 -- 1.8 
(\markcite{Colley96}Colley et al. 1996; \markcite{Tyson97}Tyson et al. 1997).
Detailed reconstructions of the multiply-lensed galaxy show a clear annular 
morphology on a scale of several kpc.  If M94 and other nearby ringed 
galaxies can be used as current-epoch analogues, the ``Pretzel Galaxy'' and 
perhaps other marginally-resolved ``core-halo'' galaxies at high redshift may 
represent youthful inner disks and bulges growing 
under the organizing influence of oval or bar 
asymmetries(\markcite{Friedli95}Friedli \& Benz 1995; 
\markcite{Waller97}Waller et al. 1997).  
Conversely, if evidence for starburst rings at high redshift 
proves to be sparse, then massive inner disks featuring ring-bar dynamics
have yet to form in most systems, or starbursting bulges are masking their
presence.

\acknowledgements 
We thank David Adler, Gene Byrd, Francoise Combes, Daniel Friedli, and 
David Westpfahl
for generously providing consultation on the dynamics of ringed-barred spiral 
galaxies.  WHW is grateful to John Huchra and the OIR division of 
the Harvard-Smithsonian Center for Astrophysics for their kind hospitality 
during his visiting appointment at the CfA.  WHW also thanks Eric Murphy and
Christine Winslow, Tufts undergraduates who helped craft some of the graphics.
UIT research is funded
through the Spacelab Office at NASA Headquarters under Project number 440-51.
We are deeply grateful to the crew of STS-67 and the many people who helped
make the {\it Astro-2} mission a success.  WHW acknowledges partial
support from NASA's Astrophysics Data Program (071-96adp).

\vfill
\eject

\figcaption{
A wide-field J-band image (extracted from the Digital Sky Survey) shows the 
outermost portions of the (R)SA(r)ab galaxy M94, 
highlighting the ovoid disk and outer stellar ring.
The field of view is $22.67' \times 22.67'$. North is up and East to the left.
Corresponding fields of the $9.66' \times 9.66'$ B-band image and the 
$7.28' \times 7.28'$ R-band, H$\alpha$ and FUV images are indicated.}

\figcaption{ 
{\bf (a.)} --- 
UV and R-band imaging of M94.  The field of view is $7.28' \times 4.53'$.
North is up and East to the left.
The UIT's FUV image shows the 
{\it starburst ring} in high contrast against a mostly dark disk.  
Exterior to the ring
are two hitherto unrecognized 500-pc size {\it bi-symmetric knots} 
on diametrically opposite sides of the nucleus.  
The HST's NUV 
image (see {\it inset}) 
shows a 450-pc long nuclear {\it mini-bar} that had been 
previously inferred
from photometric analyses of optical-band images.  By comparison, the R-band
image shows the underlying bulge and oval disk 
made up of cooler and typically older
stars.
{\bf (b.)} ---
Radial distribution of FUV intensities (surface brightnesses) reveals 
strong enhancements at the nucleus and starburst ring, along with 
regularly-spaced low-level enhancements that are associated with arcs of
FUV emission interior to the ring.}

\figcaption{
Location of IUE apertures with respect to the inner disk and nucleus of M94.
Numbers refer to the entries in {\bf Table 2}.  Positional accuracy is 
estimated at $\pm$10$''$. 
{\bf (a.)} --- Overlay of IUE's far-UV (SWP) apertures on the 
UIT/FUV image of M94, showing both nuclear and disk-dominated observations.
{\bf (b.)} --- Overlay of the IUE's near-UV (LWP and LWR) apertures on 
the UIT/FUV image of M94.}

\figcaption{
{\bf (a.)} --- The nuclear mini-bar and inner disk:
Spatial filtering of the R-band image (shown in {\bf Fig. 2}) 
shows a mini-bar with the same approximate extent and 
position angle as that seen in the HST's NUV
image.  Optical and UV spectroscopy of this region indicates a stellar
population with an early 
main-sequence turnoff (A4-A7) superposed on an older G-type population 
belonging to the central disk and bulge.  The inner disk shows dark 
spiral 
arcs of relatively lower surface brightness
that connect with the starburst ring.
{\bf (b).} --- Color morphology:  The $(R - I)$ color image is coded so that
dark features denote relatively red colors and bright features denote bluer
colors.  The image shows reddened 
arcs to the west and northeast along with especially blue knots in the 
starburst ring.}

\figcaption{
{\bf (a.)} --- Inner disk in the light of H$\alpha$.  
After scaling and subtracting the 
underlying red-continuum emission, the residual H$\alpha$ line emission shows 
concentrations in the starburst ring and in (only) one of the bi-symmetric FUV
knots.  The nucleus and innermost disk show a net deficit, due to H$\alpha$ 
{\it absorption} by the atmospheres of the B and A-type 
stars that dominate the 
light in these regions.
{\bf (b.)} --- Ratio of H$\alpha$ and FUV emission in the inner disk:  No 
radial displacement in the H$\alpha$/FUV intensity ratio is evident across the
starburst ring, contrary to outward or inward propagating starbirth 
scenarios.} 

\figcaption{
{\bf (a.)} --- Deep B-band image, taken with the 
Palomar/Hale 5-m telescope.  The 
field of view is $9.66' \times 9.66'$ with the central 2 arcmin saturated.   
{\bf (b.)} --- Complex of spiral structure.  The same B-band image 
has been 
spatially filtered to highlight the intermediate and small-scale structure.
The nuclear region was saturated and hence shows no structure.
In the disk, however, a complex spiral pattern is evident.
The starburst ring and bi-symmetric knots appear to be especially prominent 
parts of this overall spiral structure.  The knots are resolved into 
associations of hot stars, thus confirming their relative youth.}

\figcaption{
{\bf (a.)} --- Radial distribution of annular-averaged
FUV surface brightness in M94, expressed
in magnitudes.
{\bf (b.)} --- Radial distribution of the cumulative FUV flux in M94, showing 
most of the FUV light arising from the central R = 1.3 kpc.}

\figcaption{
{\bf (a.)} --- FUV spectrum of the inner disk, based on the signal-weighted 
average of 5 IUE/SWP disk-dominated spectra.
{\bf (b.)} --- NUV spectrum of the inner disk, based on the signal-weighted 
average of 2 IUE/LWP disk-dominated spectra.
{\bf (c.)} --- Optical spectrum of the central D = 8.1 arcsec, as obtained
with the Lick/Shane 1-m telescope and Image Dissecting Spectrometer (IDS).}

\figcaption{
{\bf (a.)} --- Rotation curve (radial distribution of orbital velocities), 
based on the H I observations of Mulder and van
Driel (1993), where key morphological features are noted.
{\bf (b.)} --- Radial profile of gas surface densities.  The H I gas densities
come from Mulder and van Driel (1993), and the H$_2$ are derived from the CO
observations of Gerin et al. (1991).  Beyond a radius of 75$''$, the H$_2$
surface densities have been extrapolated and most likely represent 
upper limits (see text).  This was done to provide a (probable) 
lower limit on the 
gravitational stability index (Q).
{\bf (c.)} --- Gravitational stability index (Q), as normalized by the gas
velocity dispersion V(disp) (= $\sigma$ in text).  
Beyond a radius of 75$''$, the 
extrapolated value of Q represents a (probable) lower limit.  As shown, the
starburst ring and inner disk appear 
to be situated in a shallow minimum of Q.}

\figcaption{
Radial profile of orbital frequencies and corresponding resonances.
From the H I rotation curve shown in {\bf Fig. 9a.}, the resulting 
orbital resonance diagram shows key morphological features coincident with 
important resonances.  Here, the solid, short-dashed, and long-dashed lines
respectively trace the angular frequencies $\Omega$, $\Omega \pm \kappa/4$,
and $\Omega \pm \kappa/2$.  A single pattern speed of 
35 km s$^{-1}$ kpc$^{-1}$ (where d = 4.6 Mpc and $i = 40^{\circ}$) would place
the nuclear bar inside the Inner Inner Lindblad Resonance ({\bf IILR}) 
[where $\Omega_p = \Omega - \kappa/2$], the starburst ring between the IILR \&
Outer Inner Lindblad Resonance ({\bf OILR}), the bi-symmetric knots at the 
``ultra-harmonic'' resonance ({\bf UHR}) 
[where $\Omega_p = \Omega - \kappa/4$], the 
oval disk terminating at the ``co-rotation'' resonance ({\bf CR}) 
[where $\Omega_p =
\Omega$], and the outer pseudo-ring at the Outer Lindblad Resonance ({\bf OLR})
[where $\Omega_p = \Omega + \kappa/2$].}

\begin{deluxetable}{lllllrr}
\tablewidth{40pc}
\footnotesize
\tablecaption{Log of Images (As Presented)}

\tablehead{\colhead{Telescope} & \colhead{Camera/Filter} & 
\colhead{Image \#} &  
\colhead{RA(2000)} & \colhead{Dec(2000)} & \colhead{t(exp)} & \colhead{Date} 
\\[.2ex]
\colhead{} & \colhead{} & \colhead{} & \colhead{(hr:min:sec)} & 
\colhead{($^\circ$:$'$:$''$)} & \colhead{sec} & \colhead{(d/m/y)}\\[.2ex]
\colhead{(1)} & \colhead{(2)} & \colhead{(3)} & \colhead{(4)} 
& \colhead{(5)} & \colhead{(6)} & \colhead{(7)}}
\startdata

UIT 0.38-m & FUV/B1-band & 2508 & 12:50:52.88 & 41:07:19.52 & 1040 & 
12/03/1995 \nl

HST 2.4-m & FOC/F152W & x1ar5401t & 12:50:53.04 & 41:07:12.68 & 596 & 
18/07/1993 \nl

Hale 5-m & TEK3/B-band & --- & 12:50:52.19 & 41:07:10.39 & 600 & 11/02/1994 \nl

KPNO 0.9-m & RCA-1/R-band& --- & 12:50:52.88 & 41:07:19.52 & 60 & 
18/02/1986 \nl

KPNO 0.9-m & RCA-1/I-band & --- & 12:50:52.88 & 41:07:19.52 & 100 & 
18/02/1986 \nl

KPNO 0.9-m & RCA-1/H$\alpha$-band & --- & 12:50:52.88 & 41:07:19.52 & 1055 & 
18/02/1986 \nl

Oschin 1.23-m & Schmidt/J-band & --- & 12:50:53.04 & 41:07:13.80 & 
--- & ca. 1957
\enddata

\end{deluxetable}
  
\begin{deluxetable}{llllrrrl}
\tablewidth{40pc}
\footnotesize
\tablecaption{Log of IUE Spectra}

\tablehead{\colhead{No.} & \colhead{Camera/\#} &   
\colhead{RA(2000)} & \colhead{Dec(2000)} & \colhead{P.A.} & \colhead{t(exp)} & 
\colhead{Date} & \colhead{Notes} 
\\[.2ex]
\colhead{} & \colhead{} & \colhead{(hr:min:sec)} & 
\colhead{($^\circ$:$'$:$''$)} & \colhead{deg} & \colhead{sec} & 
\colhead{(d/y)} & \colhead{}\\[.2ex]
\colhead{(1)} & \colhead{(2)} & \colhead{(3)} & \colhead{(4)} 
& \colhead{(5)} & \colhead{(6)} & \colhead{(7)} & \colhead{(8)}}
\startdata
FUV Spectra& \nl

1.& SWP15887& 12:50:52.71& 41:07:15.20& 185& 13500& 361/1981& dn \nl

2.& SWP15905& 12:50:52.71& 41:07:15.20& 183& 18000& 364/1981& dn \nl

3.& SWP28042& 12:50:51.28& 41:07:08.22& 81& 10200& 87/1986& d \nl

4.& SWP28043& 12:50:53.01& 41:07:10.24& 81& 4200& 87/1986& nd \nl

e5.& SWP28047& 12:50:54.79& 41:07:14.27& 80& 11400& 88/1986& d \nl

6.& SWP28380& 12:50:53.80& 41:07:30.22& 26& 15240& 86/1986& d \nl

7.& SWP54247& 12:50:53.50& 41:07:08.22& 81& 14700& 95/1995& nd \nl

 & \nl

NUV Spectra& \nl

1.& LWP7907& 12:50:54.79& 41:07:14.27& 80& 8400& 88/1986& d \nl

2.& LWP8289& 12:50:53.80& 41:07:30.22& 26& 9000& 146/1986& d \nl

3.& LWR12221& 12:50:52.61& 41:07:14.22& 186& 8160& 360/1981& nd \nl

4.& LWR12238& 12:50:52.71& 41:07:15.21& 182& 12000& 365/1981& nd \nl
 
\enddata

\tablecomments{(1) --- The numbers refer to the apertures shown in 
{\bf Figure 3a} (FUV) and {\bf Figure 3b} (NUV).
(5) --- Position angle of $20'' \times 10''$ aperture (in degrees).
(8) --- Notes refer to inclusion of disk (d) and/or nucleus (n) within the 
spectroscopic aperture -- with the dominant component appearing first -- as 
ascertained from visual inspection of the overlays in {\bf Figure 3.}}

\end{deluxetable}

\end{document}